# Self-assembly driven by hydrodynamic interactions in microfluidic devices.

B. Shen[1], J.Ricouvier[1], M. Reyssat[1], F. Malloggi[2], P. Tabeling[1]

## Abstract

Recent progress in colloidal science has led to elaborate self-assembled structures whose complexity raises hopes for elaborating new materials. However, the throughputs are extremely low and consequently, the chance to produce materials of industrial interest, for instance, groundbreaking optical devices, harnessing complete three-dimensional band gaps, is markedly low.

We discovered a novel hydrodynamic effect that may unlock this bottleneck. It is based on the dipolar flow interactions that build up when droplets are slowed down by the microchannel walls along which they are transported. Coupled with depletion forces, we succeeded to form, via a continuous flow process, at unprecedented speeds and under exquisite control, a rich ensemble of monodisperse planar and tridimensional clusters, such as chains, triangles, diamonds, tetahedrons, heterotrimers, possessing geometrical, chemical, and/or magnetic anisotropies enabling directional bonding. Continuous productions of millions of building blocks per second for elaborating new functional materials can be envisioned.

1 MMN, ESPCI, UMR Gulliver,10 rue Vauquelin, 75005 Paris , France
2 UMR 3299 CEA/CNRS NIMBE-LIONS, CEA Saclay, F-91191 Gif-sur Yvette, France

The field of colloidal self assembly has made a leap over the last few years by synthesizing building blocks, whose structural complexity enables the elaboration of materials possessing new, interesting properties[1,2,3,4,5,6,7,8], such as complete three dimensional (3D) photonic band gaps in the visible spectrum[2,5]. However, the evolution towards complexity has been made at the expense of the speed, and, with the approaches undertaken at the moment, it is unclear whether it will be possible, in a near future, to satisfy the high throughput conditions needed for developing practical applications. This represents a major bottleneck. In the past, colloidal materials of millimeter sizes could be produced in hours by drying colloidal solutions, but these materials relied on the assembly of identical colloids and therefore could only give rise to phases of simple symmetry, such as cubic centred, hexagonal closed paked or body centred cubic, improper for obtening interesting material functionalities[6]. In order to circumvent this difficulty, researchers have recently synthesized complex building blocks, with localized bonding sites favoring the formation of materials with non trivial crystallographic structures[8]. However, in order to assemble such building blocks – often called "colloidal molecules"[1] - , each individual colloidal « atom » must undergo a lengthy brownian search to find its right location on the colloidal "molecular" scaffold. This is too slow a process. Consistently with Ref[8], simple estimates show that thousands of seconds are needed to synthesize *one* colloidal molecule . Consequently, with an unrealistic yield of 100%, and a thousandfold parallelized process, the production of a millimeter cube material incorporating billions of such colloidal molecules would take thirty years[9]. Because of the slowness of the brownian dynamics, the impossibility to accelerate it substantially, added to moderate or mediocre yields, due to the stochastic nature of the process and the unavoidable presence of local potential minima which trap the system in long live metastable states, our ability to synthesize functional colloidal materials of industrial interest with the existing techniques can be questioned

Bringing the particles to the right locations  using a laminar flow would certainly accelerate the self assembly process by orders of magnitudes along with achieving 100% yields because of the deterministic nature of the process, but this represents a formidable challenge. The difficulty is to generate local circulations on the scale of the colloidal particles, that would transport each entity in a way suitable for synthesizing a desired structure. This raises an interesting challenge for microfluidic technology, whose goal is to precisely control flows at the micrometric scale. In fact, thus far, microfluidics has not significantly impacted the field of colloidal assembly[10,11,12,13]. Investigators have placed particles into droplets rather than in wells, in an attempt to raise the throughput but this has not been sufficient to convey significant progress. Rearranging particles with inertial forces has been demonstrated [14,15], but this approach, although developing extremely high throughputs, is not adapted for producing assemblies relevant for material science applications.  Recently, it has been shown that it is possible, theoretically,  to form arbitrary assemblies of particles in microfluidic chambers by imposing specific sequences of pressures steps at multiple fluid entries[16,17]. This concept is awaiting experimental support. In the present work, we propose a new pathway, that bears on the combined action of droplet-droplet attraction and dipolar hydrodynamical interactions, the latter being mediated by the microchannel walls along which droplets are transported  at small Reynolds numbers. The continuous flow process employed in this method allows us to produce aggregates with anisotropic structures (geometric, chemical, magnetic)at an almost 100% yield, under high throughput conditions. This approach has the potential to generate large libraries of building blocks, that could eventually be assembled to synthesize interesting functional materials of decent sizes in a realistic time frame.  The goal of the paper is to report these findings.

**Results**

**Device and flow organization**. As sketched in Fig. 1a, the microfluidic device is fabricated using a two-level lithography. It is based on a geometry published in Ref[18], in which droplets of submicrometric sizes were produced under excellent control. The geometry is thereby well adapted to operate with colloidal objects. The thinner channels of our microfluidic device (heigths between 1 and  10 µm), are located upstream. In this part, the immiscible fluids meet at the  T-junction, and, owing to a specific synchronization procedure (see Supp Mat 1), sequences of plugs, either isolated or grouped by pairs of different chemical compositions, are generated. Transported by the flow, the plugs arrive at a step marking the separation between the thin channels in which they have been generated and a deeper microchannel, called « self assembly channel » (SA channel) in which they are transported downstream.

At the step, owing to the formulations we use – we operate with surfactant concentrations far above the cmc, thereby favoring the development of adhesive depletion forces (see Materials and Methods) - , the plugs break up into sticky droplets that form aggregates (see Fig. 1a). As these clusters are transported along the SA channel, they rearrange internally, adopting eventually stationary shapes.

Two additional flow entries, located at 200 μm from the vertical symmetry plane of the device, provide an additional control on the plug break up process and the cluster dynamics in the SA channel. These additional flows contribute to confine the droplet aggregates in a narrow tube parallel to the mean flow in the SA channel (see Fig 1b) and dilute them, i.e increase the distance between two successive clusters. The effect of the control flow is shown in Supp Mat 2.

**Figure 1 : Sketch of the device, flow structure and droplet positioning in the self assembly channel.** (a) Sketch of the two-level microfluidic device. In the thinner part, located upstream, plug pairs (red and orange) are generated, and arrive at a step where they breakup into sticky droplets, forming clusters that are processed and conveyed downstream in the (deeper) SA channel. (b) Confocal image of the system, taken over long exposure times, that reveal that the clusters are confined in a tube whose center is located close to the mid plane of the SA channel and comparison with COMSOL simulations (Supp. Mat 3), that show that the flows emanating from the central inlet are confined in a tube, also located close to the SA channel symmetry axis. (c) Cluster position measurements using Adaptive Focus, for different cluster sizes. The cluster position along the "vertical" z axis (normal to the bottom wall), normalized by the SA channel height $h_2$, is shown as a function of the streamwise distance $x$ to the step normalized in the same manner. The plot shows that, within the range of experimental conditions we considered, and after a short transient, that does not exceed two $h_2$, the planar clusters self-center in the SA channel. All the experiments are performed in PDMS systems, with $h_1$=1 μm, $h_2$=10 μm and d = 5 μm using

fluorinated oil, water and 2 % SDS. Different symbols are used for N=4, to indicate that different pressure conditions at the control flow entry have been used. The ratio $h_2/d$ is thus two in these cases (d) Diamond speed $U_C$ as a function of the mean flow upstream speed $U_\infty$, obtained by dividing the flow-rate through the SA channel by the cross-sectionnal area. The full lines represent the passive advection hypothesis, in two limiting cases: $U_C = U_\infty$ (if the cluster obstructed the SA channel) and $U_C = \frac{3}{2}U_\infty$ (if the cluster was pointwise and located on the symmetry axis of the SA channel). The PDMS system dimensions are $h_1$=1 µm, $h_2$=22 µm, $w$= 20µm, and fluorinated oil in water was used with 2% SDS. The pressures are respectively 903 mbar, 861 mbar and 1-8 mbar at, respectively, the dispersed, continuous and control entries of the device of Fig 1a. Scale bar is 5µm.

**Droplet formation at the step and cluster behavior in the SA channel:** Depending on the flow conditions, plugs arriving at the step break up into droplets of identical or different sizes. There is in fact a devil staircase[19] that controls the droplet generation process, since two frequencies of emission interact with the first associated to the plug production and the second to the plug breakup at the step. In practice, the droplet sizes along with the droplet number per cluster, are precisely controlled by adjusting the flow-rates. This part is detailed in Supp Mat 1.

The droplets generated at the step are initially localized close to the top wall of the SA channel (see Fig. 1a,b). As already mentioned, owing to the sticky formulations we use[20,21] (see Materials and Methods), the droplets aggregate together just after they are generated, without coalescing. We found it useful to distinguish between two situations, depending on whether the droplet clusters tend to form a planar structure in the SA channel or whether they adopt 3D configurations therein. The former case occurs at « moderate » adhesion, i.e when the droplets stick onto each other without deforming the interfaces substantially at the contact points. 3D structures are obtained at « large adhesion », i.e when the interfaces are significantly flattened at the contact points. In both cases, the adhesion mechanism results from depletion forces developed in the presence of micelles. Large adhesion is obtained by adding salt and therefore lowering the electrostatic repulsive barrier.

The plots and images of Fig. 1 are obtained in the 2D case. Here, the aggregates tend to form planar structures that self-center in the SA channel as they travel downstream. This feature is shown by the confocal image of Fig. 1b which displays the time averaged trajectories of doublets, i.e clusters including two droplets. The long time exposed image shows that the doublets are confined in a tube. The tube is parallel to the channel walls and its center lies close to the horizontal symmetry plane of the SA channel (Fig 1b). The height of the tube does not exceed 50 µm in height, which implies that the two droplets, whose diameters are 50µm, must lie in the same horizontal plane. The structure of the cross section of the tube shown in Fig 1b suggests that the two droplets move downstream with an oblique orientation with respect to the main flow. The optical measurements of Fig. 1c obtained in a variety of flow conditions, with clusters of different sizes, confirm the centering of the clusters in the SA channel. The COMSOL study of Fig. 1b, detailed in Sup Mat 3, brings interesting information on the self centering effect. It shows that, for a single phase flow, the streamlines associated with the largest speeds, that originate from the central inlet channel, are confined in a narrow tube located approximately at the same position as the droplet clusters tube. The strong correlation between the two sub images of Fig 1b indicates that the self centering effect is kinematically-driven. Fig 1d shows that cluster speeds are significantly below those they would adopt if they were passively advected by the flow. This effect, described in the literature for the case of isolated droplets[22,23], is presumably due to the cluster friction against the channel walls. Fig 1d also indicates that in our case, the slowing down effect is significant. For a height over droplet diameter ratio of 1:3, cluster speeds are reduced between 40 and 50%, in comparison with those they would adopt if they were passively advected. Similar features are observed for 3D clusters.

**Self assembly kinetics :** The droplets form clusters that reorganize spontaneously as they travel downstream in the SA channel; this phenomenon is displayed in Movies 1-3, for clusters including two, three and four identical droplets. Here, we analyse this remarkable phenomenon in detail. In Fig. 2 b-d, obtained for moderate adhesion, we track 2D clusters located initially at 300 µm (Fig. (b),(c)), 240 µm (Fig. (d)) and 20 µm (Fig. (e)) downstream from the step, i.e in the plateau region of Fig. 1c. The microscope objective being of high numerical aperture (NA = 1.3, with a depth of field of 20 μm) and the droplet interfaces being well focused, we confirm that the structures are planar in those cases. In Fig. 2 b-c, the clusters have initially the form of bent or curved chains. These morphologies originate from the fact that droplets are produced in a row, and, just after the step, the streamlines along which they are transported are curved. In the meantime, the sudden slowing down of the flow force the droplet to jam and aggregate. After they are formed, the clusters undergo internal rearrangements and eventually adopt symmetric stationary configurations, i.e a horizontal equilateral triangle for N=3, a diamond for N=4 and a flat isosceles trapezoid for N=5 (Fig. 2 b-d). The process takes a few seconds to be completed. During the conformationnal changes, droplets roll aside each other in the horizontal plane. Eventually the clusters adopt stationary configurations for which, compared to the initial conditions, the number of internal droplet-droplet contacts, *C*, is augmented and the level of symmetry of the structure is increased. This self assembly process produces structures that do not depend on the detail of the initial conditions. Similar comments can be done for the 3D case. Fig. 2 e, obtained for large adhesion, shows the formation of a compact tetrahedron at the term of a dynamical sequence, which also takes a few seconds, within which the symmetry degree of the cluster has augmented (Movie 5 shows the phenomenon).

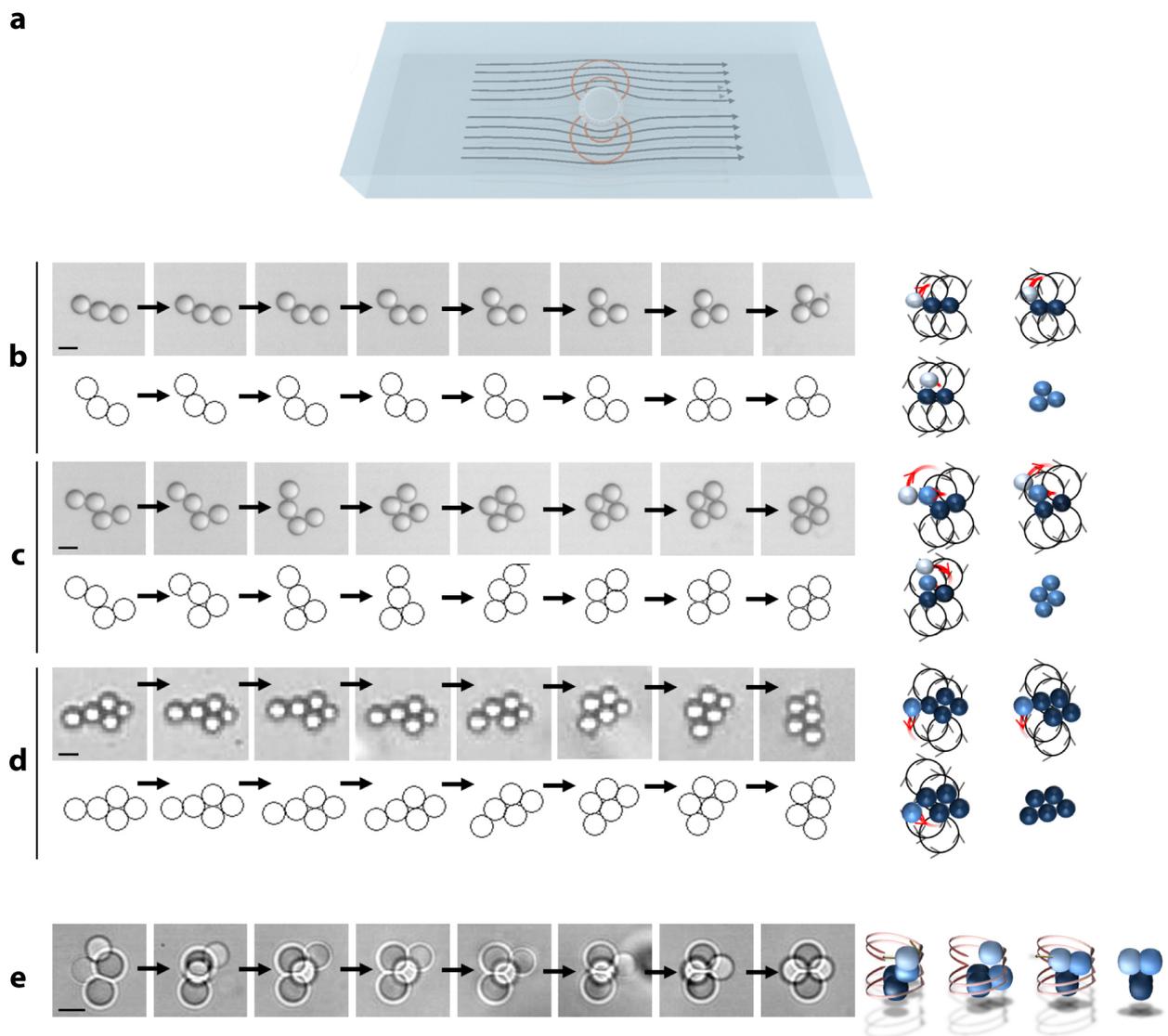

**Figure 2. Self-assembly kinetics**. (a) Schematic view of the flow field associated to an isolated droplet moving in a microchannel, at a speed smaller than the upstream flow. Blue lines correspond to the laboratory frame, orange lines represent the same flow pattern, but in the frame of reference of the upstream flow. The droplet is at the center of a dipolar recirculation system (orange lines); from this system, interactions between droplets build up. We propose that these micrometric flows are responsible for the dynamical self assembly mechanism we discover

here. (b) Dynamics of a cluster of three droplets evolving towards an equilateral triangular planar structure (see Movie 2). *Top:* Experiment made with fluorinated oil in water, 2 % SDS, with droplets 50 µm in diameter. Between each snapshot, the time interval is 0.12s. Scale bar, 50µm. *Right:* Representation of the recirculations associated to each droplet within the cluster, that bring the left droplet towards the center of the central dumbbell. Eventually, after 0.84s, the cluster reaches a steady configuration in which the effect of the recirculations cancel out. *Bottom:* numerical simulation based on Eq (1), with Y=1.5 – see the definition in the text - and time intervals of one characteristic time $\tau$ (defined in the text). (c) Same situation for N=4 (d=50 µm, time interval 0.4 s). Scale bar, 50µm (see Movie 3) Y=0.8, intervals of 6 $\tau$ and (d) N=5 (d=5 µm, time interval 1s). Scale bar, 5 µm. Y=0.8 and intervals of 15 $\tau$. (e) *Left:* Kinetics of a tridimensionnal tetrahedron, in the case N=4 (see Movie 5).. The fluids are fluorinated oil in water with 2 % SDS and 5 % Nacl. Here, d=50 µm, and the time intervals are 0.04 s. Scale bar, 5µm. *Right:* Representation of the dipolar recirculations leading to the formation of a compact tetrahedron .

The self assembly process shown in Fig. 2 b-e is not driven by brownian fluctuations because it would take days to produce the internal displacements of Fig. 2, while only a few seconds suffice in the experiments. In order to understand this self assembly mechanism, we focus on the two dimensionnal structures, which, owing to their simplicity, are more amenable to a thorough analysis. As demonstrated in Fig. 1 b-d, the planar clusters lie in a plane located close to the symmetry axis of the SA channel, i.e in a region where the shear is zero. The paradox is that they are advected by an approximately uniform steady flow and in the meantime, they develop a spectacular dynamics that leads to rapid internal rearrangements. We propose here that the physical origin of the phenomenon is linked to the presence of the top and bottom walls of the SA channel. We have found by increasing the height over droplet radius ratio, thus decreasing the confinement, that the self assembly process of Fig. 2 b-e is no longer in action. In this case, the droplets keep their initial configuration as they travel in the SA channel. This experiment indicates that the mechanism at work is mediated by the walls. The argument we propose here is thus based on the hypothesis that for moderate ratios of $h_2/d$ (where $d$ is the droplet diameter), the microchannel walls slow down the clusters' speeds, in a way shown in Fig. 1d, i.e by approximately 40-50%. The kinematic consequence of this slowing down effect is that around the droplets, a dipolar hydrodynamical field develops, as sketched in Fig 2a. In fact, in a frame moving with the average speed of the external phase, the clusters move backward, and, owing to mass conservation, the fluid that it must displace to recede recirculates in the forward direction, which gives rise to the dipolar-like pattern sketched in Fig. 2a. In fact, this reasoning can be made for each droplet embedded in a cluster. In each aggregate, the horizontal recirculations developed by each droplet exert viscous drags onto their partners, displacing them with respect to each other, therefore provoking configurational changes. The type of droplet-droplet interaction we invoke here has been studied in the literature[24,25,26,27,28,29,30], but only for the case of non sticky droplets. To the best of our knowledge, the role of dipolar interactions in aggregates has never been analyzed theoretically and the demonstration that it can play an interesting role from a self assembly prospective has never been envisaged.

The dipolar interactions we discuss here explains well, at a qualitative level, the cluster dynamics in the SA channel (see Fig. 2b-e). In the case of N=3, the droplet located at the rear of the aggregate is subjected to the recirculations generated by its partners, which work at bringing it closer to the center of the doublet they form (see Fig. 2b). As the droplet arrives at its final destination, the action of the recirculations cancels out by symmetry, and the configuration becomes stationary. In the case of N=4 the two droplets located at the rear of the structure are subjected to a similar effect (Fig. 2c). In this case, the action of the recirculations is stronger for the farther droplet, which explains that it moves more rapidly than its partner. Here too, after the droplets have reached their final destinations, forming a diamond, the dipolar hydrodynamic forces exerted onto each droplet cancel out by symmetry, and, consequently, the morphology does not evolve anymore. Similar reasoning can be done for N=5. We suggest that the same mechanism is at work in the 3D case. Fig. 2e shows how the action of horizontal, dipolar-like recirculations work at transforming a three-dimensional featureless aggregate into a symmetric closed packed tetrahedron for N=4.

**Modeling the self assembly dynamics in the case of planar structures :** The mechanism discussed above can be modelled in some details in the 2D case. Similarly as in Refs[24,29,30] we model the droplet-wall interactions by far-field pairwise dipolar interactions, noting that this approach remains acceptable at a semi-quantitative level when droplets touch each other[30]. We thus model the behaviors of our clusters by the following system of 2D dimensionless equations (see Supp Mat 4), placing ourselves in a frame of reference moving with $U_\infty$, taking $R$ (the droplet radius) as the reference scale and $\tau = R/\beta(1-\beta)U_\infty$ (in which $U_\infty$ is the speed at infinity) as the reference time:

$$\frac{d\tilde{r}_i}{d\tilde{t}} = \sum_{j \neq i} \left( \left(\frac{1}{\widetilde{r_{ij}}}\right)^2 e_\infty - 2 \left(\frac{1}{\widetilde{r_{ij}}}\right)^2 e_{ij}(e_{ij}.e_\infty) \right) + Y \sum_{j \neq i} \left( \left(\frac{1}{\widetilde{r_{ij}}-2}\right)^2 \right) e_{ij} + G_{ij} \qquad \textbf{(1)}$$

in which $\frac{d\tilde{r}_i}{d\tilde{t}}$ is the dimensionless speed of droplet $i$ (in which $\tilde{r}_i$ is its position, and $\tilde{t}$ the dimensionless time), $e_\infty$ the unit vector projected onto the mean flow at infinity, $\beta$ the reduction factor of the cluster speed (assumed to be due to friction against the walls, as discussed in the previous sections), $\widetilde{r_{ij}}$ the separation distance between droplets $i$ and $j$, $G_{ij}$ a repulsive short range term that prevents droplet interpenetration (see Supp Mat 4), and $Y$ is a dimensionless number given by the following expression:

$$Y = \frac{A}{72\pi\eta R^2 U_\infty(1-\beta)} \qquad (2)$$

in which $A$ is the constant used in the attractive part of the droplet-droplet potential (Supp Mat 4), and $\eta$ the external phase viscosity. In Equation (1), the first term of the RHS is the drift caused by the mean flow, the second the dipolar droplet-droplet interaction, the third the adhesive term and the last one a short range repulsive term that prevents interpenetration. To the best of our knowledge, the dimensionless number $Y$, which stands as the unique control parameter of the problem, is new. On physical grounds, it represents the ratio of the adhesive droplet-droplet forces over the dipolar forces. At small $Y$, adhesion is small and droplets separate out; while at large $Y$, droplets stick together permanently. Between the two regimes, there is a narrow range of Y within which a complicated dynamics develops ; its analysis stands beyond the scope of the paper.

We focus here on the large adhesion regimes, obtained, typically, for $Y$ well above 0.1. Solutions to Eq (1), obtained with the initial conditions of Fig. 2, are shown in Fig. 2b-d. The agreement between theory and experiment is remarkable. In all cases, the sequences of events calculated with Eq (1) coincide almost perfectly with the experiment. The model demonstrates that a dipolar interaction, coupled to adhesion conditions, nurtures a novel self assembly mechanism, that leads to the formation of symmetric structures. The excellent agreement between the model and the observations suggests that we have captured the physical mechanism at work in the experiments.

The model predicts that it takes a time $\tau$ to reach a stationary configuration. With the speeds and sizes at hand and a factor ß significantly smaller than unity, $\tau$ is on the order of a few seconds, which agrees well with the experiment. This estimate moreover shows that there is a potential to improve the speed of the self assembly process by increasing $U_\infty$ and shrinking the droplet sizes, which is feasible[18,32,33] with the step emulsification geometry we have intentionally chosen.

**Stationary structures :** At late times, i.e. after a few seconds, steady configurations are obtained. The planar structures we observed are displayed in Fig. 3a. Typically, chains are obtained at large control flows, and more compact structures are produced at smaller values of this parameter. As shown in Fig. 3a, the structures have different shapes - T, crosses, diamonds, trapezoids, triangles,..., which are all symmetric with respect to a mirror plane. The mirror symmetry was not preexisting in the initial clusters produced at the step. This confirms that the self assembly mechanism always works at driving the clusters from less to more symmetric states. This observation has a dynamical significance: symmetrisation is necessary for cancelling out the action of the dipolar recirculations generated by each droplet and consequently ensure configurational stationarity. The late time structures are maintained by depletion forces, and therefore they do not evolve after they are formed, as long as the surfactant concentration does not decrease. In particular, they remain stable on stopping the flow. This remark has a practical interest from an engineering prospective.

On Fig. 3a, the stationary planar structures are classified in function of the number $N$ of droplets they include and the number $C$ of droplet-droplet contacts they achieve. All the configurations are confined in a pink triangular-shaped domain: the lower boundary is the line $C = N-1$, corresponding to the structures achieving the smallest number of contacts, i.e the chain. The upper line $C = 2N-3$ corresponds to clusters maximizing the number of contacts. Inside the pink triangle, the structures adopt various shapes, in form of T or crosses. Apart from three exceptions, we have succeeded, by varying the flow conditions, to fill the space, i.e achieve all possible contact numbers. As long as N is smaller than 6, the structures are unique for a fixed pair $(N, C)$. However, similarly as in three dimensions, degeneracies are observed for N = 6. This is shown in Fig. 3a, in which three distinct morphologies are associated to the pair N=6, C=9 . All the configurations shown in the diagram are associated to well defined flow conditions and are therefore reproducible.

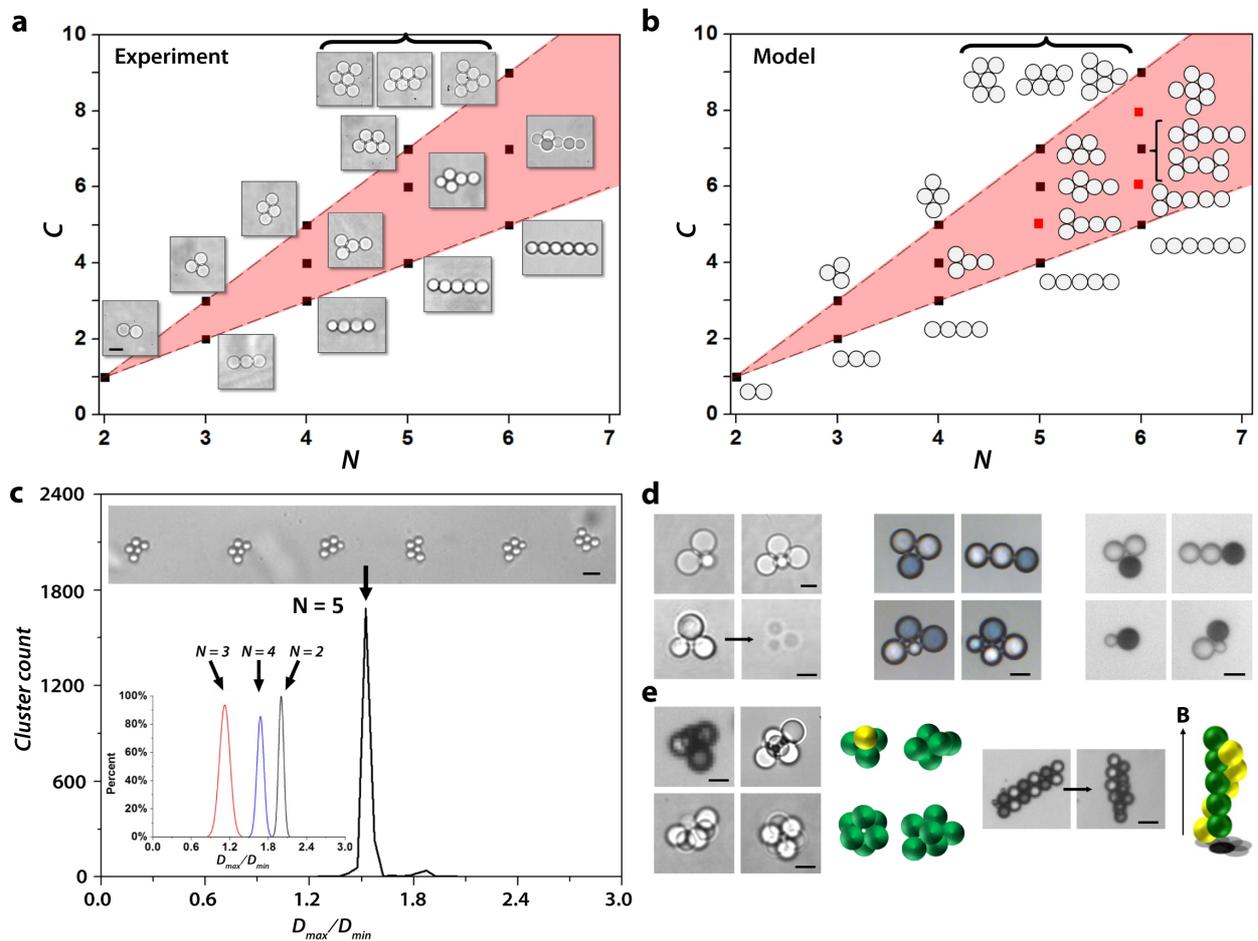

**Figure 3** : **Stationary configurations** (a) Observed cluster morphologies at late times, i.e after a few seconds. C represents the number of droplet-droplet contacts, N is the number of droplets per cluster. Scale bar is 5μm. (b) Simulation of the stationary morphologies, based on Eq (1). Red dots represent structures non observed experimentally (c) Histogram of pentamer clusters aspect ratio $\lambda = D_{max}/D_{min}$ for a population of 2134 clusters, inlcuing 5 droplets, 5μm in diameter, in which $D_{max}$ is the Feret maximum diameter and $D_{min}$ the minimum one. The coefficient of variation (i.e., the quantity 100 $\sigma/<\lambda>$) of the major cluster population is ~ 2%, where $\sigma$ is the standard deviation and $<\lambda>$ is the statistically averaged aspect ratio. On the picture, the scale bar is 10 μm. *Inset:* Distributions of the Feret aspect ratios for populations of 152 doublets, 382 triangles and 217 diamonds , all with droplet diameters of 50 μm. (d) A collection of clusters possessing anisotropic structures. *Left array:* shape anisotropies, (left) AB₂ type structure, (right) AB₃ trigonal pyramid; (Scale bar, 5μm), (bottom) Sol gel transition using water,TEOS,TEA,Minerail oil and 2 % Span 80 (scale bar 5μm) *Central array:* chemical anisotropies for N=3 (compact AB₂ and linear AB₂ type structures) and combinations of chemical and geometrical anisotropies (heterogeneous quadrimers). Scale bar, 50 μm. *Right array*, obtained with fluorinated oil in water, with 2 % SDS; Various structures with droplets incorporating ferromagnetic particles (Mickey mouse shape AB₂, linear AB₂, heterodimer, heterotrimer). On the lower figure, combination of magnetic and geometric anisotropies. Scale bar, 50 μm. (e) Three dimensional clusters. *Left array:* Polyhedron clusters with N = 4, 5, 6, and AB₃ tetrahedron structure, with a magnetic droplet (black) capable of developing a directional, localized magnetic bonding. Scale bar, 50 μm for tetrahedron structure and 5μm for the others clusters (with N = 5, 6). *Center left:* Representations of the 3D structures using yellow for the magnetic and green for the non magnetic droplets. *Center Right:* Evolution of a magnetic structure into an helix in the presence of homogeneous magnetic field whose direction is signalled by the arrow (Scale bar, 40 μm). *Right*: Representation of the 3D spiral structure, using yellow colour for the magnetic and green for the non magnetic droplets.

We used Equation (1) to calculate the ensemble of steady configurations generated by the system at late times. They are displayed in Fig. 3b. We found that all the steady structures observed experimentally are reproduced numerically. Nonetheless, Eq (1) figures out three stationary structures that the experiments have not revealed: two are in form of a T, the other is a H, and the third one has the form of a mushroom. As in the experiment, the structures are unique for a given couple *(N, C)* except for *N* = 6 where degeneracies appear. In this case, Eq (1) obtains three stationary states for *C* = 9 (as in the experiment) and two for *C* = 7 – only one was observed in this case (Fig 3a) . As a whole, the experiments and theory agree extremely well, if we consider that, in the experiments, the space of initial conditions is more difficult to span thoroughly than in a numerical model.

The notion of high throughput and close to 100% yield are demonstrated in Fig. 3c. The sharp peak shown in the Figure represents the histogram of the aspect ratios of 2134 clusters produced in a continuous way, i.e with no interruption. 95% of them have exactly the same compact trapezoidal structure, with *N=5* and *C=7* ( Fig 3c). Such structures have less than 2% geometrical dispersivity (estimated with the standard type distribution of Fig. 3c). The rest of the clusters (5%) are in form of crosses, T., or, in a few cases, transient morphologies. Comparable performances have been obtained for *N= 2, 3* and *4* (see the histograms in Fig 3c). In those cases, 100 % yields are obtained.

The hydrodynamic self-assembly mechanism we discovered here is obviously not restricted to identical droplets. This remark is crucial from the prospective of material synthesis, since, as said in the introduction, heterogeneities in the basic blocks, capable of driving directional interactions, must be present to obtain materials with interesting functionalities[4,5]. Fig. 3d shows that, with our approach, stationary structures including unlike droplets, with heterogeneous chemistries and shapes, and different magnetic properties, can be produced under full control. The phenomenon is shown in Movie 4 for the case of a triangle including a chemical heterogeneity. In fact, Fig. 3d illustrates well the classification of anisotropic structures proposed in Ref[6], based on shapes, geometry and chemistry. Structures with unlike droplets are obtained by working with plugs that break up into sequences of droplets of different sizes, according to the devil's staircase dynamics mentioned earlier[19] (see Supp Mat 1). Structures with different chemical compositions are obtained by exploiting the full capability of the devices, i.e driving plug pairs of different compositions into the SA channel (see Fig. 1a). Structures with magnetic droplets are obtained in the same way, with magnetic nanoparticles injected in one of the dispersed phases. With ferromagnetic droplets, we can explore a broader configurational space. Fig 3e shows a cluster assembling two rows of six droplets, one ferromagnetic, the other non. On applying a magnetic field, the structure undergoes a morphological change that give rises to a spiral. Also interesting from a material propective, is the fact that, on using a sol-gel transition, the clusters can be solidified.

Our approach also produces 3D clusters with high yields and large throughputs. Fig 3e displays a subset of the 3D structures we obtained. All are close packed polyhedrons, homo or heterogeneous. Despite the fact that self assembly is driven by hydrodynamic interactions *and not* by brownian fluctuations[31], the structures we obtain are unique up to *N=5* and degenerate for *N=6*, exactly as for the brownian case. This remark may suggest a different way to understand the problem of configurational selection, which is still open at the moment. Indeed, similarly as in 2D, the chemical compositions of the droplets forming these structures can be varied at will, as long as there is no interference with the surfactants. Fig. 3e shows a tetrahedron where the base is composed of ferromagnetic droplets, while the top includes a non magnetic one. The opposite situation (one ferromagnetic droplet, three non magnetic) has also been obtained (not shown). These structures have the capability to develop localized (magnetic) bonding, whose importance for material synthesis is stressed in the literature.

**Discussion and conclusion :** We have discovered a novel mechanism that assembles particles (droplets in our case), to produce an ensemble of elementary structures of great diversity. The clusters we produce can conceivably be incorporated into large assemblies to obtain a material, provided they can be stabilized and extracted from the microfluidic device. The mechanism we discovered is based on hydrodynamic dipolar interactions, which, compared to brownian motion, accelerates the self assembly process by orders of magnitude. With this mechanism, we succeeded in forming a rich ensemble of micrometric structures with controled anisotropies (geometrical, chemical, magnetic) in 2D and 3D, as well as spirals , using a continuous flow process (as opposed to the batch process used in the field at the moment),. Many of the micrometric sructures have capabilities, from a structural standpoint, to develop directional interactions. The syntheses are achieved rapidly (a few seconds) under high throughput conditions with almost one hundred percent yields. Three critical conditions that, coupled with the complexity level that is accessible, open a new pathway towards the elaboration of functional colloidal materials of sizes appropriate for industrial applications. In our case, producing $10^9$ structures in a one thousandfold parallelized device would take 30 days instead of 30 years with the standard approach. This is still a long time but there is much room for considerably improving the situation. Flow speeds can be increased by increasing pressures as long as the flow remains in the step emulsification regime[32], smaller (submicrometric) droplets can be obtained by downscaling [19,33] , and more massive parallelization can be realized. The perspective of increasing production rates, given by $\tau^{-1} = \beta(1-\beta)U_\infty R^{-1}$ by orders of magnitude, are thus excellent. Down the road, as we mentioned in the abstract, production rates of millions of clusters per second can be envisioned. Other issues that must be addressed for producing a material with our approach is the stabilisation of the cluster, their extraction from the chip and their final assembly. Stabilisation should not raise any particular difficulty, owing to the considerable range of possibilities that are avaible in terms of surfactant chemistry, photocurable materials, polymerisation techniques, etc.The solidified cluster shown in Fig. 1d indicates that solutions exist. The final assembly of the clusters can be done by drying, with or without electrical or magnetic field, and adding a few hours to the overall process. The fact that, in principle, the droplet chemisty can be changed at will, along with their physical properties (refraction index, electrical conductivity, magnetic susceptibility,…) should greatly facilitate the realization of these steps and, in the meantime, enable the production of materials with exciting functionalities.

## Materials and Methods

**Microfluidic devices.** The microfluidic devices are made by standard soft photolithography and replica-molding techniques from polydimethylsiloxane (PDMS). The molds are prepared using photolighography of a UV-curable epoxy (SU8 20XX series, Microchem). They consist of two-layers structures with different heights. The first layer includes one or several T-junctions and is followed by a shallow terrace. The depths of the thin channels ($h_1$) vary between 1 to 10 um and their widths between 10 μm to 100 μm while those of the collecting channels, i.e. SA channel ($h_2$) vary between 22 μm and 163 μm, with a width of 600 μm.

**Fluids and surfactants.** We use different formulations: direct O/W emulsions and inverte W/O emulsions. To produce oil in water structures, we use Fluorinated oil (FC3283, 3M$^{TM}$) as the dispersed phase and water with surfactant Sodium dodecyl sulfate (SDS) ( [c] varie between 0.5 CMC and 10 CMC) as the continous phase. In another case, we use deionized water as the dispersed phase and mineral oil with Span80 (2%) as the continous phase. The formulations with surfactants above the CMC develop adhesive forces between droplets, thanks to depletion forces generated by the presence of micelles. In the meantime droplet coalescence is prevented, owing to the presence of a stable film between the droplets. By adding salt in the O/W emulsion, the repulsive electrostatic barrier is lowered and adhesion between droplets is substantially enhanced. This is reflected by a flattening out of the interfaces at the droplet-droplet contact point.

**Hybrid clusters chemical composition.** Two kinds of formulations have been used to produce hybrid clusters. Methylene blue ($C_{16}H_{18}N_3SCl$), a blue ionic dye, is added at high concentration in the dispersed aqueuse phase. It yields a blue solution when dissolved in water at room temperature. The surface tension water/air is lowered from 72 mN/m to 60 mN/m at high concentrations of methylene blue. Mineral oil with 2% surfactant Span 80 is used as carrier fluid in this case. As for the preparation of the magnetic clusters, we use two different kinds of superparamagnetic particles. The first kind consists of aqueous dispersion of micro-spheres (magnetic core (maghemit $Y - Fe_2O3$) and silica matrix), with thiol group grafted to their surfaces (purchased from Chemicell GmbH). The size of particles is around 500nm. The second type of colloidal particles consists of magnetite $Fe_3O_4$ nanoparticles of 40nm which is dispersed already in octane (80%) (purchased from Ademtech) and which we can further dilute with mineral oil to obtain 0.01 %v/v of ferrofluid in mineral oil. The solution is stabilized by 2% Span80 in the organic phase We use mineral oil with 2% Span 80 as carrier fluids in the previous aqueous dispersion and water with 2% SDS for the latter.

**Solidifiable clusters.** Silica sol solution is achieved by mixing 2 ml tetraethyl orthosilicate (TEOS, 99.0%, Fluka) and 0.2ml triethyl amine (TEA, 99.5%, Fluka) with 10 ml water under stiring at room temperature until one single homogeneous phase appeared ($\sim$ 8h). The precursor aqueous solution is used as dispersed phase and mineral oil with 2% surfactant Span 80 as the continuous and dilution phase. We use PDMS system with W=20μm, $h_1$ = 1μm, $h_2$ = 22μm to produce droplets of 5 μm diameter and develop a sol-gel transition leading to cluster solification in the fluid at rest.

**Fluid driving and measurement equipement.** To drive the fluids, we use pressure sources (MFCS Fluigent) or syringe pumps NEMESYS. By using an integrated flowmeter in the case of pressure sources, we could measure the flow-rates of the external phase injected in the different entries. Throughout the experiments, we span a range of flow rates varying between 5 and 100μl/min. The droplet motions are recorded with a fast camera (Photron) through an inverted microscope (Zeiss or Leica). Image processing is used to determine the droplets characteristics.

**Adaptive focus z position measurement.** By using the fully automated Leica microscope system with Adaptive Focus Control (AFC), the measurement of the vertical position (z coordinate) of the clusters could be performed. These measurements were made on a microfluidic device with $w$ = 20μm, $h_1$ = 1μm and $h_2$ = 22μm. After the detection of the lower wall z = 0 is performed, we span the height of the microfluidic SA channel ($h_2$ = 22μm), up to the upper wall, plot intensity profiles and localize the maximum to determine the cluster "altitude" $z$. The process being reproducible , averaging over many clusters was carried out.

**Microscope confocale imaging.** We produced a stationary train of clusters made of aqueous droplets in mineral oil. To improve the quality of visualization, the dispersed phase was mixed with fluorescein isothiocyanate dextran. Rhodamine B red dye ($6 \times 10 - 3$% in aqueous solution), was infused into the channels and washed before the experiments. This dye permeates the PDMS matrix. Fluids were pumped into the devices through PEEK tubing using a pressure controller (MFCS Fluident). The microfluidic system was characterized by $w$ = 50μm, $h_1$ = 10μm and $h_2$ = 163μm. A series of experiments with different flow conditions was performed, with pressures at the control entries varying from 250 mbar to 700 mbar. Before of a slow exposure time, clusters could not be resolved at the individual level; their averaged trajectories form a florescent tube on Fig 1a.

**Acknowledgements**
We gratefully acknowledge Ministry of Research, ESPCI and CNRS for their support of this work. We thank B. Cabane, A. Leshansky, R.Dreyfus, D.Pines, S.Quake, C.Ren for enlightening discusions, F. Monti for technical help, A. Simon for help in the preparation of the manuscript, all the MMN for fruitful interactions.

**Author contributions**
B.S made almost all the experiments; J.R made experiments for the magnetic case; P.T, J.R. wrote the model and solved it numerically; F.M. made the COMSOL simulation. F.M.,B.S.,J.R., M.R., made the confocal study; P.T., B.S.,J.R.,F.M.,M.R. wrote the paper.

# Figure Caption

**Figure 1 : Sketch of the device, flow structure and droplet positioning in the self assembly channel**. (a) Sketch of the two-level microfluidic device. In the thinner part, located upstream, plug pairs (red and orange) are generated, and arrive at a step where they breakup into sticky droplets, forming clusters that are processed and conveyed downstream in the (deeper) SA channel. (b) Confocal image of the system, taken over long exposure times, that reveal that the clusters are confined in a tube whose center is located close to the mid plane of the SA channel and comparison with COMSOL simulations (Supp. Mat 3), that show that  the flows emanating from the central inlet are confined in a tube, also located close to the SA channel symmetry axis. (c) Cluster position measurements using Adaptive Focus, for different cluster sizes. The cluster position along the "vertical" z axis (normal to the bottom wall), normalized by the SA channel height $h_2$, is shown as a function of the streamwise distance $x$ to the step normalized in the same manner. The plot shows that, within the range of experimental conditions we considered, and after a short transient, that does not exceed two $h_2$, the planar clusters self-center in the SA channel. All the experiments are performed in PDMS systems, with $h_1$=1 µm, $h_2$=10 µm and d = 5 µm using fluorinated oil, water and 2 % SDS. Different symbols are used for N=4, to indicate that different pressure conditions at the control flow entry have been used. The ratio $h_2$/d is thus two in these cases (d) Diamond speed $U_C$ as a function of the mean flow upstream speed $U_\infty$, obtained by dividing the flow-rate through the SA channel by the cross-sectionnal area. The full lines represent the passive advection hypothesis, in two limiting cases: $U_C = U_\infty$ (if the cluster obstructed the SA channel) and $U_C = \frac{3}{2} U_\infty$ (if the cluster was pointwise and located on the symmetry axis of the SA channel). The PDMS system dimensions are $h_1$=1 µm, $h_2$=22 µm, $w$= 20µm, and fluorinated oil in water was used with 2% SDS. The pressures are respectively 903 mbar, 861 mbar and 1-8 mbar at, respectively, the dispersed, continuous and control entries of the device of Fig 1a.  Scale bar is 5µm.

**Figure 2. Self-assembly kinetics**.  (a) Schematic view of the flow field associated to an isolated droplet moving in a microchannel, at a speed smaller than the upstream flow. Blue lines correspond to the laboratory frame, orange lines represent the same flow pattern, but in the frame of reference of the upstream flow. The droplet is at the center of a dipolar recirculation system (orange lines); from this system, interactions between droplets build up. We propose that these micrometric flows are responsible for the dynamical self assembly mechanism we discover here. (b) Dynamics of a cluster of three droplets evolving towards an equilateral  triangular planar structure (see Movie 2). *Top:*  Experiment made with  fluorinated oil in water, 2 %  SDS, with droplets 50 µm in diameter. Between each snapshot, the time interval is 0.12s. Scale bar, 50µm. *Right*: Representation of the recirculations associated to each droplet within the cluster, that bring the left droplet towards the center of the central dumbbell. Eventually, after 0.84s, the cluster reaches a steady configuration in which the effect of the recirculations cancel out. *Bottom:* numerical simulation based on Eq (1), with Y=1.5 – see the definition in the text - and time intervals of one characteristic time τ (defined in the text). (c) Same situation for N=4 (d=50 µm, time interval 0.4 s). Scale bar, 50µm (see Movie 3) Y=0.8, intervals of 6 τ   and (d) N=5 (d=5 µm, time interval 1s). Scale bar, 5 µm. Y=0.8 and intervals of 15 τ.  (e) *Left:* Kinetics of a tridimensionnal tetrahedron, in the case N=4 (see Movie 5)..  The fluids are fluorinated oil in water with 2 % SDS and 5 % Nacl. Here, d=50 µm, and the time intervals are 0.04 s. Scale bar, 5µm.  *Right:*  Representation of the dipolar recirculations leading to the formation of a compact tetrahedron .

**Figure 3** : **Stationary configurations** (a) Observed cluster morphologies at late times, i.e after a few seconds. C represents the number of droplet-droplet contacts, N is the number of droplets per cluster.  Scale bar is 5µm. (b) Simulation of the stationary morphologies, based on Eq (1). Red dots represent structures non observed experimentally (c) Histogram of pentamer clusters aspect ratio λ =$D_{max}$/$D_{min}$ for a population of 2134 clusters, inslucing 5 droplets, 5µm in diameter, in which $D_{max}$ is the Feret maximum diameter and $D_{min}$  the minimum one. The coefficient of variation (i.e., the quantity 100 $\sigma$/<λ>) of the major cluster population is ~ 2%, where σ is the standard deviation and <λ> is the statistically averaged aspect ratio. On the picture, the scale bar is 10 µm.  *Inset:* Distributions of the Feret aspect ratios for populations of 152 doublets, 382 triangles and 217 diamonds , all with droplet diameters of  50 µm. (d) A collection of clusters possessing anisotropic structures. *Left array:* shape anisotropies, (left) $AB_2$ type structure, (right) $AB_3$ trigonal pyramid; (Scale bar, 5µm), (bottom) Sol gel transition using water,TEOS,TEA,Minerail oil and 2 % Span 80 (scale bar  5µm) *Central array:* chemical anisotropies for N=3 (compact $AB_2$ and linear $AB_2$ type structures) and combinations of chemical and geometrical anisotropies

(heterogeneous quadrimers). Scale bar, 50 µm. *Right array*, obtained with fluorinated oil in water, with 2 % SDS; Various structures with droplets incorporating ferromagnetic particles (Mickey mouse shape $AB_2$, linear $AB_2$, heterodimer, heterotrimer). On the lower figure, combination of magnetic and geometric anisotropies. Scale bar, 50 µm. (e) Three dimensional clusters. *Left array:* Polyhedron clusters with N = 4, 5, 6, and $AB_3$ tetrahedron structure, with a magnetic droplet (black) capable of developing a directional, localized magnetic bonding. Scale bar, 50 µm for tetrahedron structure and 5µm for the others clusters (with N = 5, 6). *Center left:* Representations of the 3D structures using yellow for the magnetic and green for the non magnetic droplets. *Center Right:* Evolution of a magnetic structure into an helix in the presence of homogeneous magnetic field whose direction is signalled by the arrow (Scale bar, 40 µm). *Right:* Representation of the 3D spiral structure, using yellow colour for the magnetic and green for the non magnetic droplets.

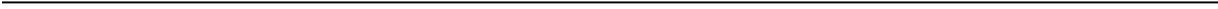